\documentclass[aps,prl,reprint,twocolumn,superscriptaddress,floatfix,nofootinbib,longbibliography]{revtex4-1}
\usepackage{graphicx,amsmath,amsfonts,amssymb,amsthm,xr}
\usepackage{epsfig,amsmath,amssymb,color,dsfont,upgreek,physics}
\usepackage{mathrsfs}
\usepackage{mathtools}
\usepackage{bbold}
\usepackage{comment}
\usepackage{float}

\usepackage[bookmarks=true,colorlinks,linkcolor=OrangeRed,urlcolor=NavyBlue,citecolor=RoyalBlue]{hyperref}
\usepackage[dvipsnames]{xcolor}
\usepackage{orcidlink}

\definecolor{mygold}{rgb}{0.93,0.69,0.13}

\definecolor{mypurple}{rgb}{0.49,0.18,0.56}

\definecolor{mygreen}{rgb}{0,0.5,0}

\definecolor{mygreen}{rgb}{0,0.5,0}
\definecolor{myred}{rgb}{0.7,0,0}
\definecolor{myblue}{rgb}{0,0,1}

\begin{document}
\title{Disorder-Free Localization in $2+1$D Lattice Gauge Theories with Dynamical Matter}
\author{Jesse Osborne${}^{\orcidlink{0000-0003-0415-0690}}$}
\affiliation{School of Mathematics and Physics, The University of Queensland, St. Lucia, QLD 4072, Australia}
\author{Ian P.~McCulloch${}^{\orcidlink{0000-0002-8983-6327}}$}
\affiliation{School of Mathematics and Physics, The University of Queensland, St. Lucia, QLD 4072, Australia}
\author{Jad C.~Halimeh${}^{\orcidlink{0000-0002-0659-7990}}$}
\email{jad.halimeh@physik.lmu.de}
\affiliation{Department of Physics and Arnold Sommerfeld Center for Theoretical Physics (ASC), Ludwig-Maximilians-Universit\"at M\"unchen, Theresienstra\ss e 37, D-80333 M\"unchen, Germany}
\affiliation{Munich Center for Quantum Science and Technology (MCQST), Schellingstra\ss e 4, D-80799 M\"unchen, Germany}

\begin{abstract}
Disorder-free localization (DFL) has been established as a mechanism of strong ergodicity breaking in $1+1$D lattice gauge theories (LGTs) with dynamical matter for quenches starting in homogeneous initial states that span an extensive number of gauge superselection sectors. Nevertheless, the fate of DFL in $2+1$D in the presence of dynamical matter has hitherto remained an open question of great interest in light of the instability of quenched-disorder many-body localization in higher spatial dimensions. Using infinite matrix product state calculations, we show that DFL survives in $2+1$D LGTs, albeit it is generally less pronounced than in $1+1$D, and highly depends on the matter configuration of the initial state. Through suitable matter configurations, we are able to relate and compare the $1+1$D and $2+1$D cases, showing that the main ingredient for the strength of DFL in our setup is the \textit{propagation directionality} of matter. Our results suggest that, generically, DFL is weakened with increasing spatial dimension, although it can be made independent of the latter by minimizing the propagation directionality of matter in the initial state.
\end{abstract}

\date{\today}
\maketitle

\textbf{\textit{Introduction.---}}Understanding and controlling the far-from-equilibrium physics of quantum many-body models has been at the forefront of condensed matter research over the last decades \cite{Eisert2015,Rigol_review,Buca2023unified}. Even though it is well known by now that interacting models with quenched disorder may lead to many-body localization (MBL) \cite{Basko2006,Gornyi2005,Nandkishore_review,Abanin_review,Alet_review}, which has also been demonstrated experimentally \cite{Schreiber2015,Choi2016}, generic interacting many-body systems are expected to thermalize based on the eigenstate thermalization hypothesis \cite{Deutsch1991,Srednicki1994,Rigol_2008,Deutsch_review}. Nevertheless, recent ergodicity-breaking mechanisms in homogeneous nonintegrable models have proven that there is more than meets the eye to this dichotomy. For example, weak ergodicity breaking can occur in the disorder-free nonintegrable PXP model, where persistent oscillations arise in the quench dynamics of local observables after starting in initial states carrying a large overlap with athermal eigenstates known as quantum many-body scars, which possess anomalously low bipartite entanglement entropy and are roughly equally separated in energy \cite{Bernien2017,Turner2018,Moudgalya2018}. An example of strong ergodicity breaking in disorder-free nonintegrable models is \textit{Stark} MBL, which arises in a chain of interacting fermions subjected to a sufficiently strong constant electric field \cite{vanNieuwenburg2019,Schulz2019}, with recent experimental realizations showcasing this \cite{Scherg2020,Morong2021}.

Another recent paradigm of strong ergodicity breaking that has garnered a lot of attention is disorder-free localization \cite{Smith2017,Brenes2018}. DFL has been particularly prominent in gauge theories, which are quantum many-body systems hosting local (gauge) symmetries that enforce an intrinsic \textit{locally constrained} relationship between matter and the surrounding gauge field \cite{Weinberg_book,Gattringer_book,Zee_book}. In quantum electrodynamics (QED), the associated $\mathrm{U}(1)$ gauge symmetry manifests in the famous Gauss's law. Local constraints play a crucial role in various fields of physics, from quantum spin liquids \cite{Balents_NatureReview,Gingras2014} and anyonic quantum computing \cite{Kitaev2003} to fractonic models \cite{Sala2020,Khemani2020}. In gauge theories, they can also conspire to bring about DFL after quenches of special initial states prepared as a superposition \cite{Smith2017,Brenes2018} or thermal ensemble \cite{Halimeh2022TDFL} over an extensive number of gauge superselection sectors. It is important to note that DFL can occur even when these states are themselves translation-invariant and the quench Hamiltonian is disorder-free and interacting. DFL has been theoretically demonstrated in various models \cite{smith2017absence,Metavitsiadis2017,Smith2018,Russomanno2020,Papaefstathiou2020,McClarty2020,hart2021logarithmic,Zhu2021,Sous2021}, and methods have also been proposed to stabilize it against gauge-breaking errors in modern quantum simulators \cite{Halimeh2021stabilizingDFL,Halimeh2021enhancing,Lang2022stark}.

Although works on DFL have mostly focused on $1+1$D gauge theories with dynamical matter, works studying DFL in $2+1$D have been few and only considered pure LGTs, i.e., without dynamical matter \cite{karpov2021disorder,Chakraborty2022}. Even though one can in principle employ Gauss's law to integrate out matter degrees of freedom in, e.g., $\mathrm{U}(N)$ and $\mathrm{SU}(N)$ LGTs, this usually comes at the cost of extending the range of interactions, introducing additional local constraints, and breaking the gauge symmetry \cite{Zohar2019removing}, which gives rise to a model vastly different from the pure LGT. Indeed, it is known that retaining dynamical fermionic matter in $2+1$D can lead to significantly richer physics, such as, e.g., possible exotic quantum spin liquid phases in regularized lattice QED \cite{Hashizume2022}. Furthermore, recent concrete proposals for the realization of different kinds of $2+1$D LGTs with dynamical matter on cold-atom platforms \cite{Homeier2022quantum,Osborne2022,Surace2023Abinitio} provides further impetus to understand DFL behavior in such models.

In this work, we consider a regularized model of $2+1$D QED with dynamical matter, and numerically calculate the quench dynamics of various matter configurations prepared either in the physical sector of Gauss's law, or as a superposition over an extensive number of gauge superselection sectors. We find ergodic behavior in the former case, while the latter leads to DFL in the form of a finite imbalance plateau at late evolution times. We analyze the strength of DFL in $2+1$D between the different matter configurations, and also compare to the $1+1$D case. We find that the \textit{propagation directionality} of the initial matter configuration plays the most important role in determining how strong DFL is. We expect our conclusions to be applicable to other kinds of LGTs with dynamical matter hosting different gauge groups and also in higher spatial dimensions.

\textbf{\textit{Model.---}}A useful regularization of lattice QED is the so-called \textit{quantum link formulation}, which entails replacing the infinite-dimensional gauge and electric-field operators with spin-$S$ operators \cite{Chandrasekharan1997,Wiese_review,Kasper2017}. This facilitates numerical and experimental implementations while still capturing the wealth of physical phenomena of lattice QED. In the following, we will consider the spin-$1/2$ $\mathrm{U}(1)$ quantum link model (QLM), given by the Hamiltonian
\begin{align}\nonumber
    \hat{H}=\sum_\mathbf{r}\bigg[&-\kappa\sum_{\nu=x,y}c_{\mathbf{r},\mathbf{e}_\nu}\Big(\hat{\phi}_\mathbf{r}^\dagger\hat{s}^+_{\mathbf{r},\mathbf{e}_\nu}\hat{\phi}_{\mathbf{r}+\mathbf{e}_\nu}+\text{H.c.}\Big)\\\label{eq:QLM}   &+mc_\mathbf{r}\hat{\phi}^\dagger_\mathbf{r}\hat{\phi}_\mathbf{r}-J\Big(\hat{U}_{\square_\mathbf{r}}+\hat{U}^\dagger_{\square_\mathbf{r}}\Big)\bigg],
\end{align}
where $\mathbf{r}=\big(r_x,r_y\big)^\intercal$ is a vector denoting the location of a site on the $2$D lattice, $\mathbf{e}_\nu$ is the unit vector along the direction $\nu\in\{x,y\}$, and the tunneling constant $\kappa=1$ sets the overall energy scale. The tunneling and mass terms of Hamiltonian~\eqref{eq:QLM} are staggered as per the Kogut--Susskind formulation \cite{Kogut1975}, where $c_{\mathbf{r},\mathbf{e}_x}=1$, $c_{\mathbf{r},\mathbf{e}_y}=(-1)^{r_x}$, and $c_\mathbf{r}{=}(-1)^{r_x{+}r_y}$. The lattice spacing is set to unity throughout. The fermionic annihilation and creation operators $\hat{\phi}_\mathbf{r}$ and $\hat{\phi}_\mathbf{r}^\dagger$, respectively, satisfy the canonical anticommutation relations $\big\{\hat{\phi}_\mathbf{r},\hat{\phi}_{\mathbf{r}'}\big\}=0$ and $\big\{\hat{\phi}_\mathbf{r},\hat{\phi}_{\mathbf{r}'}^\dagger\big\}=\delta_{\mathbf{r},\mathbf{r}'}$. They act on the matter field at site $\mathbf{r}$ with $m$ denoting the fermionic mass. At the link between sites $\mathbf{r}$ and $\mathbf{r}{+}\mathbf{e}_\nu$, the gauge and electric fields are represented by the spin-$1/2$ operators $\hat{s}^\pm_{\mathbf{r},\mathbf{e}_\nu}$ and $\hat{s}^z_{\mathbf{r},\mathbf{e}_\nu}$, respectively. Magnetic interactions for the gauge fields are governed by the plaquette term $\hat{U}_{\square_\mathbf{r}}=\hat{s}^+_{\mathbf{r},\mathbf{e}_x}\hat{s}^+_{\mathbf{r}+\mathbf{e}_x,\mathbf{e}_y}\hat{s}^-_{\mathbf{r}+\mathbf{e}_y,\mathbf{e}_x}\hat{s}^-_{\mathbf{r},\mathbf{e}_y}$ at strength $J$.

The $\mathrm{U}(1)$ gauge symmetry of Hamiltonian~\eqref{eq:QLM} is generated by the local operator
\begin{align}\label{eq:Gr}    \hat{G}_\mathbf{r}=\hat{\phi}^\dagger_\mathbf{r}\hat{\phi}_\mathbf{r}{-}\frac{1{-}(-1)^{r_x+r_y}}{2}{-}\sum_{\nu=x,y}\Big(\hat{s}^z_{\mathbf{r},\mathbf{e}_\nu}{-}\hat{s}^z_{\mathbf{r}-\mathbf{e}_\nu,\mathbf{e}_\nu}\Big),
\end{align}
which can be viewed as a discretized version of Gauss's law. 

\begin{figure}[t!]
	\centering
	\includegraphics[width=\linewidth]{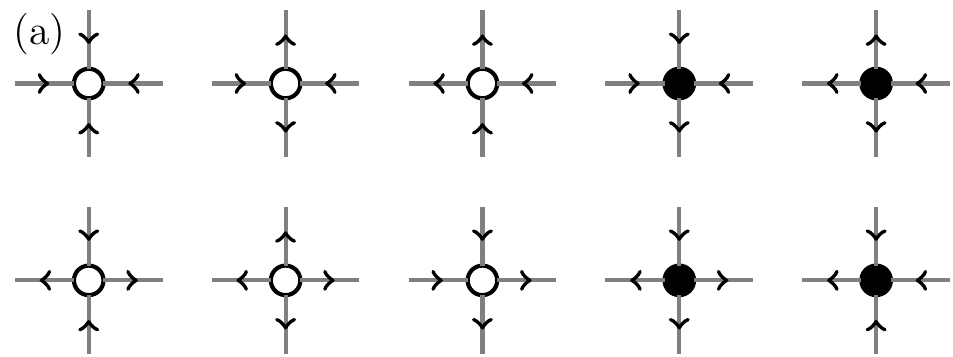}
	\includegraphics[width=\linewidth]{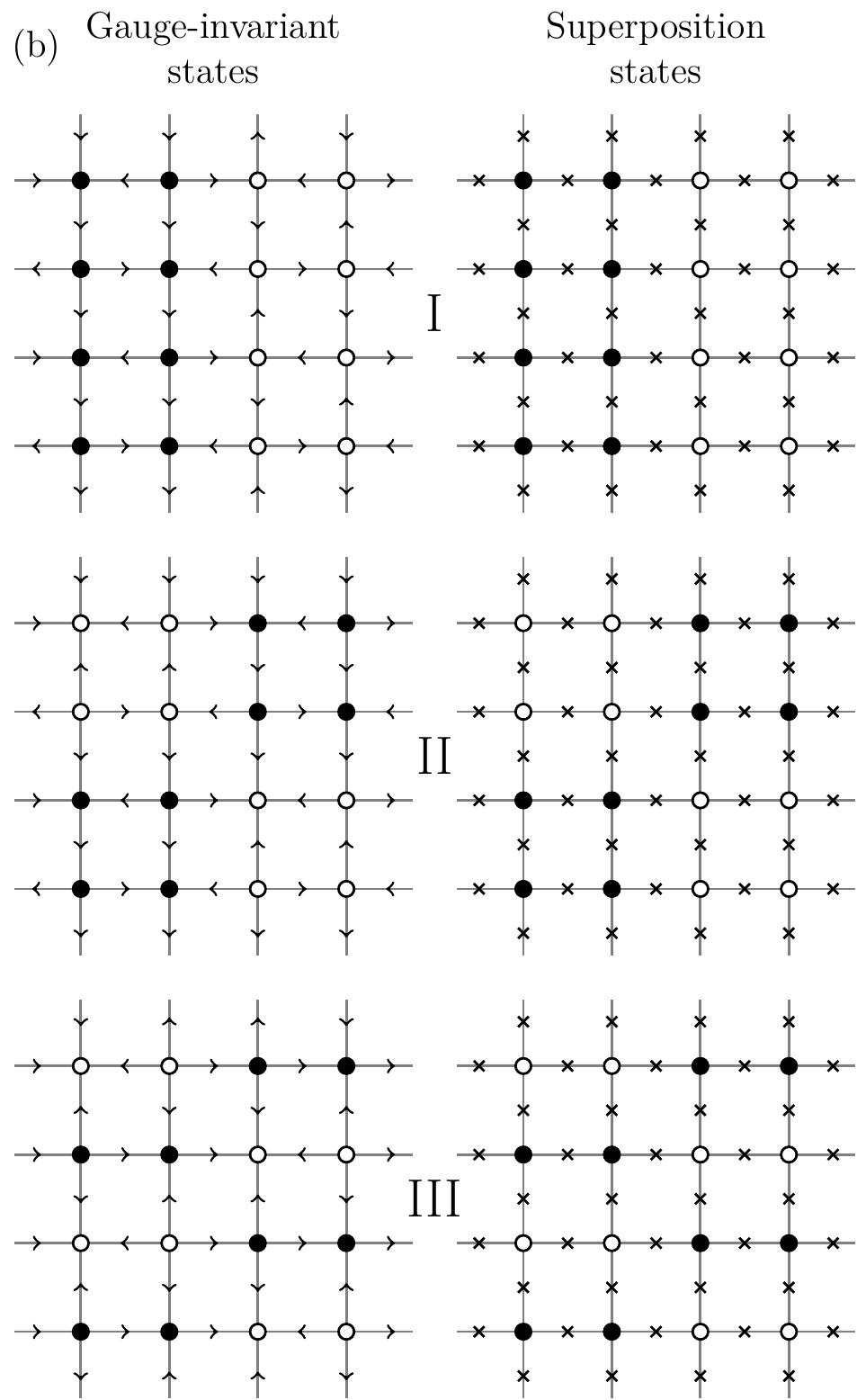}
	\caption{(a) The configurations at each site and its four neighboring links as allowed by Gauss's law~\eqref{eq:Grph} without background charges, i.e., $g_\mathbf{r}=0,\,\forall\mathbf{r}$. (b) The initial states considered in this work comprise three matter configurations: I, II, and III, all of which are at half-filling. The left column shows these configurations in a gauge-invariant state within the physical sector: $\hat{G}_\mathbf{r}\ket{\psi_0}=0,\,\forall\mathbf{r}$. The arrows on links denote eigenstates of the local $\hat{s}^z$ operator. The right column shows these configurations in a superposition over an extensive number of gauge superselection sectors. The crossmarks on the links denote the ground state of the local $\hat{s}^x$ operator.}
	\label{fig:schematic}
\end{figure}

We now perform the particle-hole transformation \cite{Hauke2013}
\begin{subequations}
\begin{align}
    &\hat{\phi}_\mathbf{r}\to\frac{1+(-1)^{r_x+r_y}}{2}\hat{\phi}_\mathbf{r}+\frac{1-(-1)^{r_x+r_y}}{2}\hat{\phi}_\mathbf{r}^\dagger,\\
    &\hat{s}^+_{\mathbf{r},\mathbf{e}_\nu}\to(-1)^{r_x}\hat{s}^x_{\mathbf{r},\mathbf{e}_\nu}-i(-1)^{r_y}\hat{s}^y_{\mathbf{r},\mathbf{e}_\nu},
\end{align}
\end{subequations}
in order to get rid of the Kogut--Susskind staggering in Eq.~\eqref{eq:QLM}. This renders the Hamiltonian in the experimentally friendly form \cite{Osborne2022}
\begin{align}\nonumber
    \hat{H}=\sum_\mathbf{r}\bigg[&-\kappa\sum_{\nu=x,y}\Big(\hat{\phi}_\mathbf{r}\hat{s}^+_{\mathbf{r},\mathbf{e}_\nu}\hat{\phi}_{\mathbf{r}+\mathbf{e}_\nu}+\text{H.c.}\Big)+m\hat{\phi}^\dagger_\mathbf{r}\hat{\phi}_\mathbf{r}\\\label{eq:QLMph}   &+J\Big(\hat{s}^+_{\mathbf{r},\mathbf{e}_x}\hat{s}^-_{\mathbf{r}+\mathbf{e}_x,\mathbf{e}_y}\hat{s}^+_{\mathbf{r}+\mathbf{e}_y,\mathbf{e}_x}\hat{s}^-_{\mathbf{r},\mathbf{e}_y}+\text{H.c.}\Big)\bigg],
\end{align}
where now the generator reads
\begin{align}\label{eq:Grph}
    \hat{G}_\mathbf{r}=(-1)^{r_x+r_y}\bigg[\hat{\phi}_\mathbf{r}^\dagger\hat{\phi}_\mathbf{r}+\sum_{\nu=x,y}\Big(\hat{s}^z_{\mathbf{r},\mathbf{e}_\nu}+\hat{s}^z_{\mathbf{r}-\mathbf{e}_\nu,\mathbf{e}_\nu}\Big)\bigg].
\end{align}
Gauge-invariant states $\ket{\psi}$ are simultaneous eigenstates of all local generators: $\hat{G}_\mathbf{r}\ket{\psi}=g_\mathbf{r}\ket{\psi},\,\forall\mathbf{r}$. A set of these eigenvalues $g_\mathbf{r}$---commonly referred to as background charges---over the whole lattice defines a unique gauge superselection sector. In a given sector, these local constraints enforce a set of allowed configurations of the electric fields on the links adjacent to a site, depending on the matter occupation of that site; see Fig.~\ref{fig:schematic}(a).

\textbf{\textit{Quench dynamics.---}}We are interested in initial states at half filling with a finite imbalance in the matter configuration, which we then subject to a quench under Hamiltonian~\eqref{eq:QLMph} and subsequently calculate the time evolution of the imbalance. The latter is defined as 
\begin{align}
    \mathcal{I}(t)= \frac{1}{2L_xL_y} \Big\langle \psi(t) \Big| \sum_{\mathbf{r} \in S} \hat{\phi}_\mathbf{r}^\dagger \hat{\phi}_\mathbf{r} - \sum_{\mathbf{r} \notin S} \hat{\phi}_\mathbf{r}^\dagger \hat{\phi}_\mathbf{r} \Big| \psi(t) \Big\rangle,
\end{align}
where $S$ is the set of initially occupied sites, $\ket{\psi(t)}=e^{-i\hat{H}t}\ket{\psi_0}$,
and $\ket{\psi_0}$ is the initial state. We set $m=0.5\kappa$ and $J=0$, although our results are general and do not depend on these particular values, as we shall see later.
We obtain the time-evolved states $\ket{\psi(t)}$ from the initial product states, shown in Fig.~\ref{fig:schematic}(b), by using infinite matrix product state (iMPS) techniques~\cite{Uli_review,Paeckel_review,mptoolkit}: specifically, we use a version of the time-dependent variational principle (TDVP) algorithm~\cite{Haegeman2011,Haegeman2016} adapted to work in the thermodynamic limit.
We use the standard procedure of representing a 2D system with a cylindrical geometry (i.e., periodic boundary conditions are employed), using a width of $L_y=4$ sites (and, therefore, also $L_y=4$ links), while the system is in the thermodynamic limit in the $x$ direction (as the states considered are invariant under translations of four matter sites in the $x$ direction).
We use a single-site evolution scheme with adaptive bond dimension growth~\cite[App.~B]{vumps}, and use a time-step of $0.01/\kappa$, which achieves convergence for our most computationally stringent results over the investigated evolution times.

\begin{figure}[t!]
	\centering
	\includegraphics[width=\linewidth]{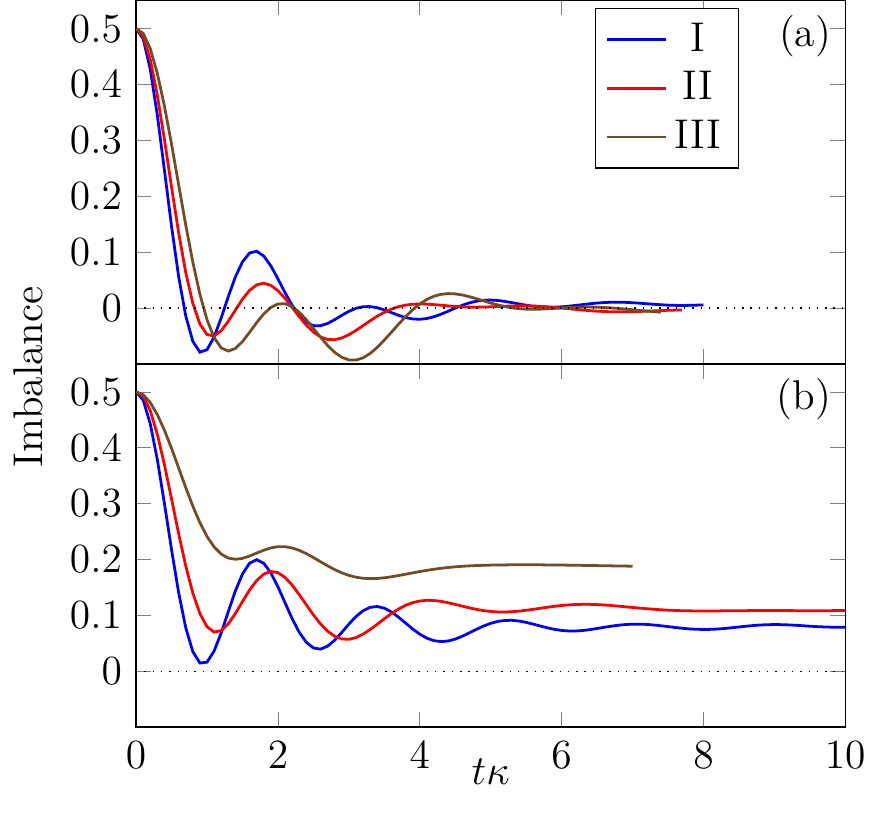}
	\caption{(Color online). (a) When the matter configurations I, II, and III are prepared in the physical sector of Gauss's law, as shown in the left column of Fig.~\ref{fig:schematic}(b), the imbalance quickly vanishes in the wake of a quench by Hamiltonian~\eqref{eq:QLMph}, indicating thermalization. (b) When the matter configurations I, II, and III are prepared in a superposition over an extensive number of superselection sectors, as shown in the right column of Fig.~\ref{fig:schematic}(b), the imbalance does not completely vanish at late times, and the system retains memory of its initial state, indicating DFL. In this case we find the strongest DFL for configuration III, followed by II, and then I. This is directly connected to their propagation directionality number $\eta=1,2,3$, respectively. DFL is weakened as $\eta$ increases (see text). For these results we have set $m=0.5\kappa$ and $J=0$, although we have checked that our conclusions are unaltered for different values of these parameters. In our MPS calculations we have used $L_y=4$ sites.}
	\label{fig:dynamics}
\end{figure}

We consider the matter configurations shown in Fig.~\ref{fig:schematic}(b). If we prepare each in a homogeneous superselection sector, say, the physical sector of Gauss's law $g_\mathbf{r}=0,\,\forall\mathbf{r}$, by preparing the local electric field on each link in an appropriate eigenstate of the local $\hat{s}^z$ spin basis, we find that upon quenching the corresponding initial state the imbalance will decay to zero at late times, as shown in Fig.~\ref{fig:dynamics}(a). This is indeed not surprising, since Hamiltonian~\eqref{eq:QLMph} is translation-invariant, hosts no disorder, and is nonintegrable, and therefore thermalization is expected and does occur.

However, upon preparing the electric field on each link in the ground state of the local $\hat{s}^x$ spin basis, see right column of Fig.~\ref{fig:schematic}(b), each matter configuration forms a state that is a superposition over an extensive number of superselection sectors. In $1+1$D LGTs with dynamical matter, this is known to lead to a finite value of the imbalance at late times and the absence of thermalization, i.e., DFL arises \cite{Smith2017,Brenes2018}. We find that this behavior persists in $2+1$D, as shown in Fig.~\ref{fig:dynamics}(b) for the considered matter configurations when they are in such a superposition state. 

Interestingly, we see a significant difference in the degree of DFL between the matter configuration III and its counterparts in I and II. The late-time imbalance plateau is significantly larger in this configuration. We attribute this to its low \textit{propagation directionality} compared to configurations I and II. Indeed, looking carefully at configuration III, we find that a given particle can only move to its right or left along the $x$ direction in a single-order tunneling process and not at all in the $y$ direction---according to Hamiltonian~\eqref{eq:QLMph}, tunneling can occur between neighboring sites that are either both occupied or empty. We can assign a propagation directionality number $\eta=1$ to this state, where the maximum value that $\eta$ can take in $2+1$D for a square lattice is $\eta_\text{max}=4$. On the other hand, configurations I and II have $\eta=3$ and $2$, respectively, with configuration II having a slightly larger value of the imbalance plateau than configuration I. From Fig.~\ref{fig:dynamics}(b), we therefore see that the larger $\eta$ is, the weaker DFL appears to be.

\begin{figure}[t!]
	\centering
	\includegraphics[width=\linewidth]{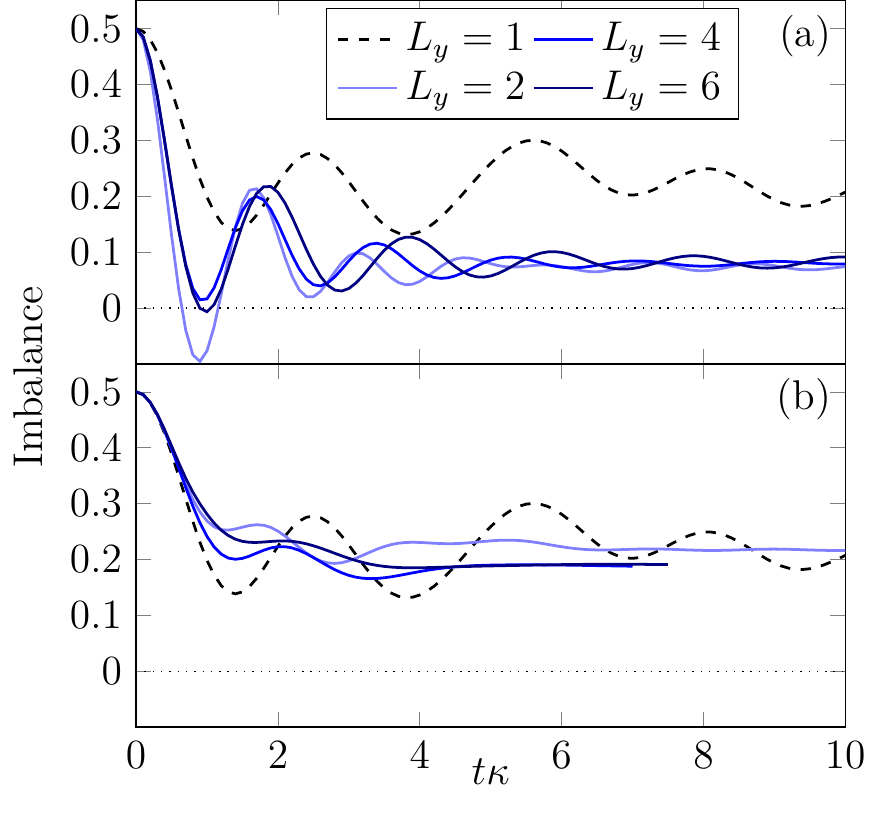}
	\caption{(Color online). Finite-size scaling analysis to connect the $1+1$D and $2+1$D cases, and to emphasize the importance of propagation directionality. (a) Configuration I in a superposition state will lead to DFL, as established in Fig.~\ref{fig:dynamics}(b) for $L_y=4$. For $L_y>1$, configuration I has a propagation directionality number $\eta=3$ while for $L_y=1$ it has $\eta=1$. Correspondingly, we find more pronounced DFL at $L_y=1$ than for $L_y>1$. Furthermore, the degree of DFL seems to be roughly the same for any value of $L_y>1$. (b) Configuration III has $\eta=1$ for all values of $L_y$, and this manifests in the same strength of DFL occurring for all considered values of $L_y$.}
	\label{fig:FSS}
\end{figure}

To better understand this behavior, we perform a finite-size scaling analysis for configurations I and III in the superposition state, the results of which are shown in Fig.~\ref{fig:FSS}. For both these configurations, our results can be directly compared to the $1+1$D case. Focusing first on configuration I in Fig.~\ref{fig:FSS}(a), we find that the $1+1$D limit of $L_y=1$ shows a significantly larger imbalance at late times than for larger $L_y$. This makes sense because for $L_y=1$ the propagation directionality of configuration I drops to $\eta=1$, whereas for $L_y\geq2$ it is $\eta=3$. This means that in $1+1$D the tunneling of matter is further constrained, leading to more localized dynamics. It is interesting to note that although there is significant difference in the degree of DFL between $L_y=1$ and $L_y>1$, there is little qualitative difference between different values of $L_y>1$. This further indicates that $\eta$ is more important than spatial dimensionality when it comes to DFL. We further cement this picture by turning our attention to configuration III, shown in Fig.~\ref{fig:FSS}(b), where we find that there is little qualitative difference in the degree of DFL between the $1+1$D case of $L_y=1$ and larger values of $L_y$. This is due to the fact that configuration III has $\eta=1$ regardless of the value of $L_y$, thereby explaining why DFL exhibits the same qualitative robustness in $1+1$D and $2+1$D.

\begin{figure}[t!]
	\centering
	\includegraphics[width=\linewidth]{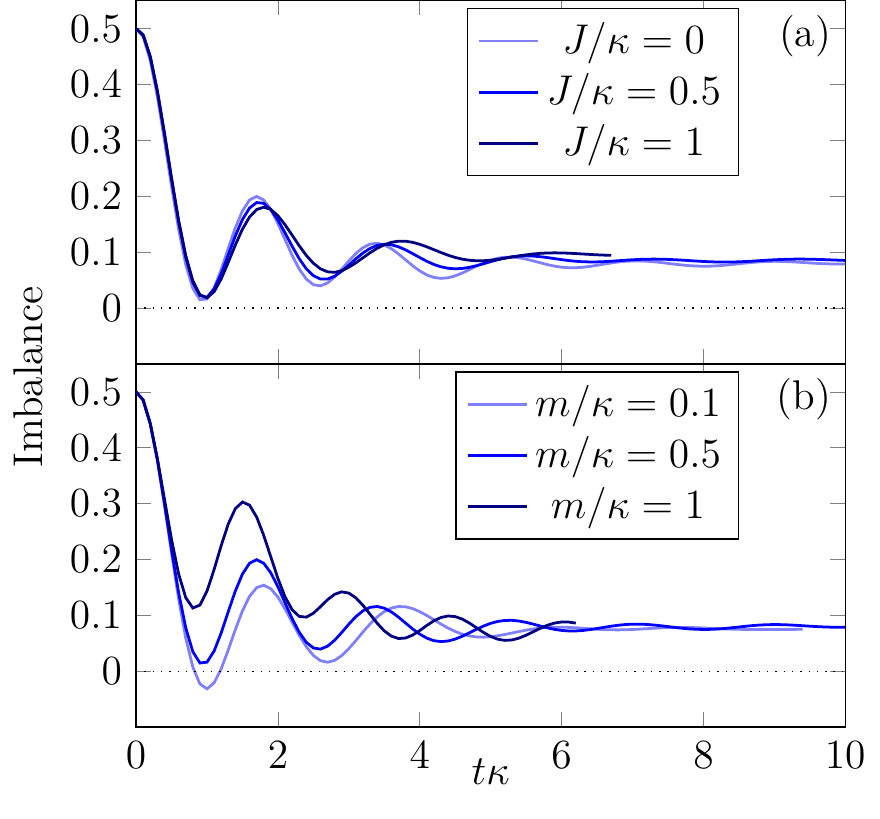}
	\caption{(Color online). Our results are general and we have checked that they remain qualitatively the same for different values of (a) $J$ and (b) $m$.}
	\label{fig:plaquette}
\end{figure}

\textbf{\textit{Discussion and outlook.---}}Using infinite matrix product state techniques, we have studied disorder-free localization in the $2+1$D spin-$1/2$ $\mathrm{U}(1)$ quantum link model with dynamical fermionic matter, a paradigmatic system of significant experimental interest \cite{Osborne2022,Surace2023Abinitio}. We have found that disorder-free localization persists in $2+1$D for initial states comprising a superposition over an extensive number of gauge superselection sectors, whereas initial states in a single homogeneous sector will lead to a vanishing imbalance in the long-time limit.

Interestingly, spatial dimensionality seems to play a less direct role in the degree of localization compared to what we term the \textit{propagation directionality} of the initial state, defined as the number of directions in which matter is allowed to propagate. The larger this number, the less localized is the late-time dynamics. Generically, this number is expected to increase with spatial dimension, which indicates that disorder-free localization will generically get weaker in higher dimensions. However, we have shown that states exhibiting a propagation directionality number of unity in $2+1$D will still exhibit robust localization, with the late-time imbalance plateau reaching roughly the same value as in $1+1$D.

We emphasize that our results are general in that we reach the same qualitative conclusions for different values of $J$ and $m$, as shown in Fig.~\ref{fig:plaquette}, and we expect our findings to apply to other lattice gauge theories in $2+1$D, such as $\mathbb{Z}_2$ lattice gauge theories and spin-$S$ $\mathrm{U}(1)$ quantum link models. Regarding $\mathbb{Z}_2$ lattice gauge theories, it would be interesting to explore the effect of adding a local-pseudogenerator term, which in $1+1$D was shown to enhance localization due to the dynamical emergence of an upgraded local symmetry containing the $\mathbb{Z}_2$ gauge symmetry. When it comes to spin-$S$ $\mathrm{U}(1)$ quantum link models, it is important to note that for $S>1/2$, the gauge-coupling term $\propto g^2\sum_{\mathbf{r},\nu}\big(\hat{s}^z_{\mathbf{r},\nu}\big)^2$ will not be an inconsequential energetic constant as in the case of $S=1/2$. By integrating out the gauge fields in $1+1$D in the Kogut--Susskind limit $S\to\infty$, $g$ was shown to serve as a tunable parameter equivalent to disorder strength \cite{Brenes2018}, and so it would be interesting to analyze its effect in $2+1$D. However, this is not expected to change the conclusions of our work in any nontrivial way, since the propagation directionality will still have the same effect on matter (de)localization. Another interesting venue to explore would be the addition of a Rokhsar--Kivelson potential \cite{Rokhsar1988} and analyze its effect on localization with dynamical matter, noting that this has been analyzed in the case of a pure (without dynamical matter) spin-$1/2$ $\mathrm{U}(1)$ quantum link model \cite{karpov2021disorder,Chakraborty2022}. It would also be interesting to test our conclusions for initial states prepared in a thermal ensemble, as was recently done in the $1+1$D case \cite{Halimeh2022TDFL}, and determine if the propagation directionality will still play a dominant role in the degree of DFL. We leave these exciting avenues open for future work.
\bigskip
\begin{acknowledgments} 
J.C.H.~acknowledges stimulating discussions with Pablo Sala, and is grateful to Fabian Grusdt, Philipp Hauke, Lukas Homeier, Johannes Knolle, and Bing Yang on related projects. I.P.M.~acknowledges support from the Australian Research Council (ARC) Discovery Project Grants No.~DP190101515 and DP200103760. J.C.H.~acknowledges funding from the European Research Council (ERC) under the European Union’s Horizon 2020 research and innovation programm (Grant Agreement no 948141) — ERC Starting Grant SimUcQuam, and by the Deutsche Forschungsgemeinschaft (DFG, German Research Foundation) under Germany's Excellence Strategy -- EXC-2111 -- 390814868. Numerical simulations were performed on The University of Queensland's School of Mathematics and Physics Core Computing Facility ``getafix''.
\end{acknowledgments}

\bibliography{biblio}
\end{document}